# Focal-ratio-degradation (FRD) mitigation in a multimode fiber link using mode-selective photonic lanterns

Aurélien Benoît,[1] Stephanos Yerolatsitis,[2] Kerrianne Harrington,[2] Tim A. Birks,[2] and Robert R. Thomson[1]

[1]SUPA, Institute of Photonics and Quantum Sciences, Heriot-Watt University, Edinburgh, EH14 4AS, UK
[2]Department of Physics, University of Bath, Claverton Down, Bath BA2 7AY, UK
*Corresponding author: a.benoit@hw.ac.uk

We present a new way to mitigate focal-ratio degradation (FRD) when using optical fibers to transport multimode light. Our approach exploits a custom multicore fiber (MCF) with six dissimilar cores that are single mode at ~1550 nm wavelength and minimally coupled over 7 m. We fabricated adiabatic mode-selective photonic lanterns (PLs) at each end of the MCF to create a fiber link with multimode ports, the PLs coupling each spatial mode of the multimode ports to a specific core of the MCF and vice versa. The PL-MCF-PL link exhibits superior FRD behavior compared to a conventional multimode fiber that also supports 6 modes, because it inhibits the transfer of light from lower-order modes to higher-order modes. These results open up a potentially powerful new approach to mitigate FRD in multimode fiber links, with particular applications in astronomical instruments.

## 1. INTRODUCTION

Optical fibers provide a powerful way to route light with low loss and are one of the most ubiquitous photonic technologies. As such, they have become enabling in areas such as telecommunications [1], laser-based industrial manufacturing [2], biomedical instrumentation [3] and physical / chemical environmental sensors [4]. One area that has also been transformed by low-loss fiber is optical- and near-infrared astronomical instrumentation, where fibers can be used to efficiently and flexibly route light from the telescope focal plane to a spectrograph for wavelength-resolved analysis [5]. In such applications, fibers facilitate the complete mechanical de-coupling of the spectrograph from the telescope. As a result, the spectrograph can be placed in an environmentally stabilized room in the observatory, maximizing opto-mechanical stability [6]. The fiber link is therefore a key component in instruments such as HARPS that are capable of measuring spectral shifts in stellar spectra on the order of 1 part in $10^{10}$ between 380 nm and 690 nm, which corresponds to detecting a radial velocity shift due to a planetary companion of ~10 cm.s$^{-1}$ [7].

Maximizing the optomechanical stability of instruments is not the only advantage that fibers can provide. In multi-object and diverse-field spectrographs, fibers can route light from many different parts of the telescope focal plane to the spectrograph. The ability of fibers to provide high "multiplex gain" in this manner has been exploited in numerous astronomical instruments. A notable example included 2dF [8], where hundreds of multimode fibers (MMFs) were used to route light from selectable regions across the focal plane to the AAOmega spectrograph for simultaneous analysis. Highly multiplexed observations by 2dF have provided remarkable insights into the large-scale structure of the universe, such as the existence of baryon acoustic oscillations [8].

Despite the numerous advantages of fibers, they are not without drawbacks. One example is focal-ratio degradation (FRD) [9-10], where light at the output of a fiber emerges with a lower f-number (higher numerical aperture (NA)) than the injected light. For photon-starved astronomical applications, the input f-ratio is generally controlled to maximize telescope-to-fiber coupling efficiency. However, if the f-ratio is then lowered by a fiber with FRD feeding an instrument downstream, the resulting "faster" beam of light requires larger instrument optics to deliver the same efficiency and spectroscopic resolution. This results in more expensive instruments and lower quality optics [11]. Techniques that efficiently mitigate FRD in fiber-fed spectrographs are therefore of considerable interest, and MMF links that are immune to FRD would have profound implications on the design of future systems.

From the perspective of guided-wave optics, FRD is the result of mode coupling along the fiber due to micro-bending or core deformations. This mode coupling can transfer light from lower-order modes to higher-order modes [12-13]. Since higher-order modes emerge from the output of the fiber with a lower f-ratio than lower-order modes, mode-coupling induces FRD. For a photonic point of view, the f-ratio is directly linked to the numerical aperture NA with the relation $f/\# = (2.NA)^{-1}$. For all the f-ratio evolutions in this paper, we present the NA value on a second axis. Figure 1 illustrates how the f-ratio of light emerging from a fiber might vary with input f-ratio, when using either an ideal MMF without FRD (green line) or a real MMF with FRD (red lines). In the ideal case, the output f-ratio matches the input f-ratio between two limits - the upper limit being the f-ratio of the fundamental mode of the fiber, and the lower limit being the f-ratio of the highest-order guided mode. In real MMFs, however, the output f-ratio is reduced compared to the input f-ratio, with the difference becoming particularly pronounced as the input f-ratio increases. Any FRD mitigation technology is aimed at achieving a fiber link with FRD properties closer to the ideal characteristic shown in Fig. 1, while still maintaining throughout efficiency.

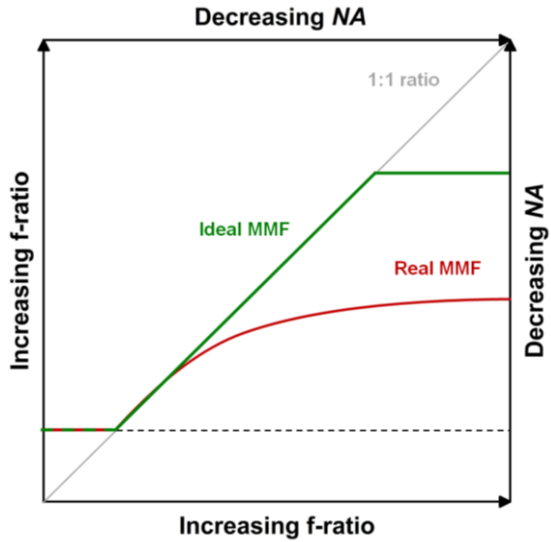

**Fig. 1**. Representation of how FRD would manifest in ideal (green) and real (red) MMF. The solid gray line represents the line along which no FRD would be generated.

The FRD behavior of MMFs has been intensively studied. The work has focused on understanding the impact of the MMF core diameter [9, 14-15], core geometry [16], wavelength [17-18], bending stress [12] and MMF length [19], and has also been extended to the case of bundles of MMFs [20-21]. In all of these cases, the MMFs follow the trend represented by the red curve in Fig. 1.

Over the last decade, a guided wave technology known as the photonic lantern (PL) has been developed which has the potential to transform the operation of MMF-fed astronomical instruments by efficiently interfacing multimode astronomical signals with single-mode (SM) photonic devices. The basic idea is shown in Fig. 2, where a gradual optical transition connects the spatial modes of an MMF core to a set of discrete single-mode cores [22-23]. In general, the connection is one-to-many: light contained in one spatial mode of the MMF is distributed between most or all of the single-mode cores, and vice versa. By designing the PL correctly, it conserves the number of guided modes in both directions [22, 24, 26]. Several astrophotonics applications for PLs have been proposed, including in OH-line suppression [27], mode scramblers [28], and for reformatting multimode light for diffraction-limited spectroscopy [24, 29-30].

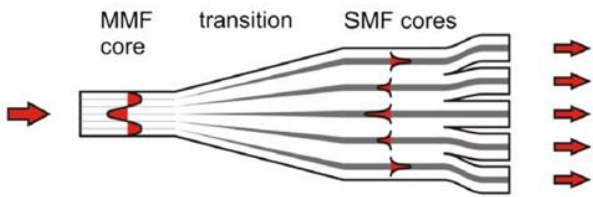

**Fig. 2.** Schematic diagram of a PL made by tapering a bundle of single-mode fibers. From right to left: separate SMFs are fused together to form a single glass body, which simultaneously reduces in cross-sectional scale to form a MMF core. Surrounding it is a low-index jacket (not shown) that forms a cladding for the MMF core [22, 26].

Here, we demonstrate the feasibility of a new route to mitigate FRD in multimode fibre links using mode-selective PLs [25]. In contrast to conventional PLs, in a mode-selective PL the connection between modes and cores is one-to-one: light in one mode of the MMF is routed to one specific single-mode core (a different one for each mode), and vice-versa. Our proposed MMF link is a multicore fiber (MCF) with mode-selective PLs at either end. If the MCF is designed to avoid inter-core coupling along its length, the whole PL-MCF-PL link preserves the power distribution across the modes between input and output, preventing FRD.

The paper is structured as follows. We characterize the 6-core MCF in Section 2 and the mode-selective PLs made from it in Section 3. In Section 4, we demonstrate the PL-MCF-PL multimode fiber link. In Section 5, we present the conclusions of the paper.

## 2. THE MULTICORE FIBER (MCF)

Our MCF had 6 Ge-doped step-index cores with dissimilar diameters of 11, 10.3, 9.5, 8.3, 7.3 and 6.5 μm (Fig. 3). Their minimum separation was 26 μm, which simulations indicated would ensure crosstalk remained below 0.08 % after 2 m of MCF [31]. To study its guiding properties, we coupled light of wavelength $1550 \pm 20$ nm from a thermal source into a 1 m length of a single-mode telecoms fiber (SMF-28). We excited each MCF core individually using direct fiber-to-fiber butt excitation, with fiber alignment optimized for maximum transmission. Near-field images of the MCF output taken with an InGaAs camera, Figs. 4(a-f), demonstrate that there is negligible cross-talk between the cores over 1.5 m. Figure 4(g) is a composite of Fig. 4(a-f), and Fig. 4(h) is a similar composite but with the fibers slightly misaligned by a lateral offset. The light patterns were unaffected by the input misalignment, demonstrating the single-mode nature of each core at 1550 nm.

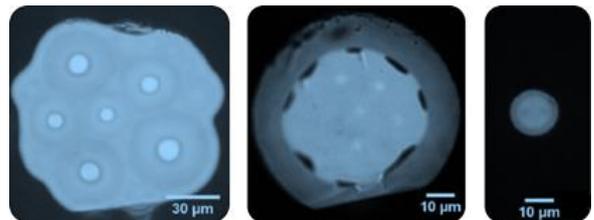

**Fig. 3**. Optical micrographs of (Left) the MCF, (Center) the MCF with an F-doped silica jacket collapsed onto it part-way along the transition, and (Right) the multimode port of a mode-selective PL.

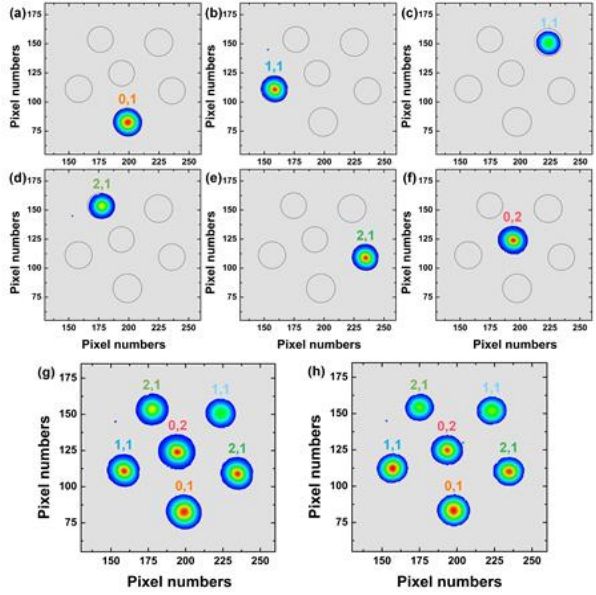

**Fig. 4.** (a-f) Near-field images of the output in each MCF core using light at around 1550 nm. (g & h) Composite images of the outputs. In (g), the excitation is optimized to maximize power transmission. In (h) the excitation is deliberately misaligned in an attempt to excite any higher order modes supported by each core.

## 3. THE PHOTONIC LANTERN (PL) TRANSITIONS

The optimum MCF core arrangement was investigated by simulating the evolution of the effective index of the guided modes as a function of the taper ratio ($\rho$) for the first 7 modes of the MCF, where the 7th mode is the lowest-order cladding mode. This was repeated for all 720 possible permutations of cores over the same set of 6 locations, the aim being to equalize the effective index separation of the modes along the transition and thus minimize mode coupling; the optimum is shown in Fig. 5. An MCF was then fabricated with this design [31].

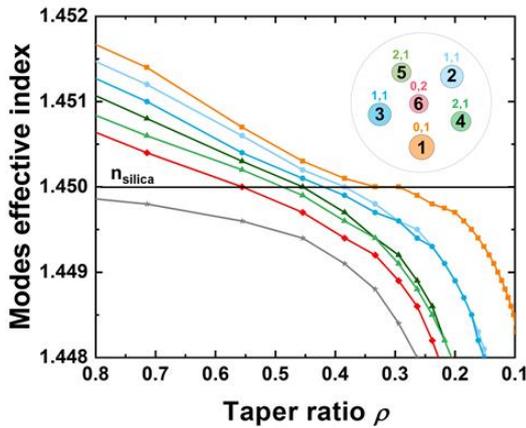

**Fig. 5.** Evolution of the mode effective index as a function of the taper ratio $\rho$ for the first 7 modes of the multicore fiber calculated by the finite-element method. Inset: schematic representation of the cross section of the MCF with the associated color pattern of each mode. The cores are numbered 1-6 in decreasing order of size, and increasing order of the mode excited in the PL. The pairs of numbers indicate the LP mode designation - for example, core 2 excites an $LP_{1,1}$ mode.

A PL was fabricated at one end of 7 m of MCF using established tapering techniques [25]. The profile of the taper was approximately linear with a length of 3 cm. The cleaved end of the taper had a multimode core with a diameter of 12 μm and an NA of 0.22 [31]. To investigate the mode-selective properties of the PL transition we used the system described in Section 2. Light at $1550 \pm 20$ nm was coupled into each MCF core individually, and the output of the PL was imaged, Fig. 6. Each MCF core excites a specific LP mode at the MMF port of the PL - clear evidence of the mode-selective nature of the PL and the minimally coupled cores at 1550 nm.

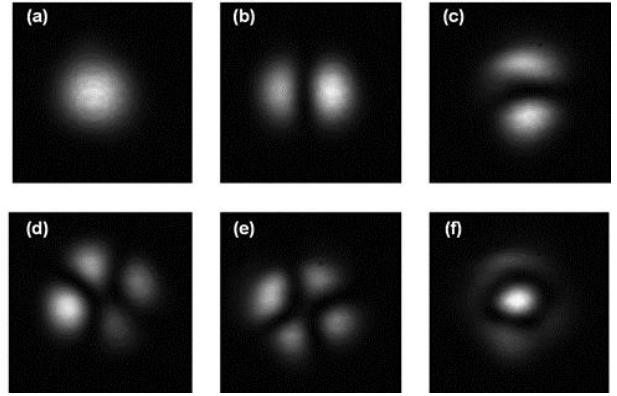

**Fig. 6.** Near-field images of the PL output while coupling 1550 nm light into the opposite end of the 7 m long MCF. (a), (b), (c), (d), (e) and (f) were obtained while individually exciting the MCF cores labelled 1, 2, 3, 4, 5, and 6 respectively in Fig. 5. The patterns closely resemble the $LP_{0,1}$, $LP_{1,1}$, $LP_{1,1}$, $LP_{2,1}$, $LP_{2,1}$ and $LP_{0,2}$ modes respectively.

To confirm the suitability of the PL transition for the FRD-mitigation application, we investigated its mode coupling properties using the experiment shown in Fig. 7(a). A thermal tungsten light source (Thorlabs SLS201L/M) was used to generate broadband light from the visible to the near infrared. This light was coupled into a 400 μm diameter fiber which was used to transmit the light to the characterization system. The light emerging from this fiber was collimated to a diameter of 8 mm by lens L1, and coupled through a bandpass filter to select the wavelength band for the measurement. An adjustable graduated pinhole was used to restrict the diameter of the collimated beam, and thus the f-ratio of the light focused by lens L2 onto the MMF port of the PL. The output of the MCF was imaged onto an InGaAs camera to view the distribution of the light across its cores. Figure 8 summaries the evolution of the relative power contained in each core as a function of the excitation f-ratio and (inset) the MCF output measured under the lowest f-ratio excitation (unlike Fig. 4, this is not a composite image.). At high f-ratio excitation, only the fundamental mode (orange squares) is excited in the multimode port of the PL, and as a consequence light only emerges from the bottom right MCF core. As the input f-ratio reduces, higher-order modes are increasingly excited in the multimode port and light becomes increasingly distributed across all the cores of the MCF. In effect, the PL is functioning as a mode analyzer. All the different steps of the evolution of the output light core distribution versus input f-ratio are presented in the Supplementary Material in Fig. S1 and are animated in Visualization 1.

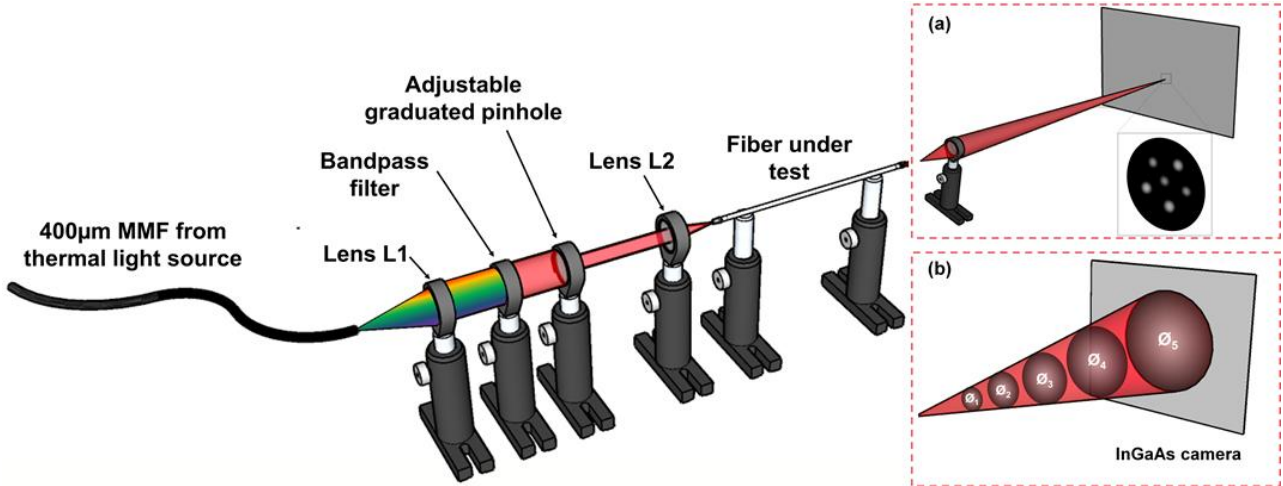

**Fig. 7.** Schematic of the experiment used to characterize the guiding and FRD properties of our fiber devices. Light from the thermal light source was coupled into a 400 μm core fiber. The light emitted by this fiber was collimated by lens L1 and passed through a bandpass filter to control the wavelength of the characterization. The filtered light was then passed through an adjustable graduated pinhole to control its diameter and coupled into lens L2 to excite the fiber under test. The f-ratio was controlled via the adjustable pinhole. At the output of the fiber we could either use system (a) to characterize the near-field intensity distribution, or system (b) to characterize the f-ratio of the output. In both cases, light detection is performed using an InGaAs camera.

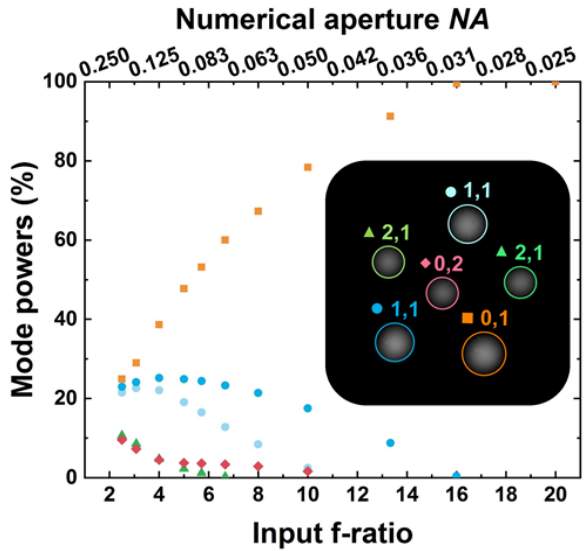

**Fig. 8**. Evolution of the relative power of each mode at the output of the MCF as a function of the f-ratio of the light used to excite the PL. Inset: A near-field image of the MCF output when the PL is excited using low f-ratio light. Each core is labelled to indicate the LP mode to which it is coupled at the input of the PL. The evolution is presented in Visualization 1 and is described in Fig. S1 in the Supplementary Material.

## 4. THE FRD-RESISTANT MULTIMODE FIBER LINK

The full FRD-resistant PL-MCF-PL link was realized by fabricating mode-selective PL transitions at both ends of the 7 m MCF. FRD performance was characterized using the system in Fig. 7(b). Figure 9 (red circles) presents the output f-ratio of the PL-MCF-PL link as a function of the input f-ratio, the latter being the ratio of the focal length of L2 to the diameter of the adjustable graduated pinhole. These results were obtained using $1550 \pm 20$ nm light, the wavelength for which the MCF cores were designed to be single mode and minimally coupled. To compare its FRD to a conventional multimode fiber, we also fabricated a step-index Ge-doped fiber that supported 6 modes at 1550 nm (6-MMF). The 6-MMF had a core diameter of 10.4 μm, an 80 μm diameter cladding and a numerical aperture of 0.22. Both fiber devices were held as straight as possible over the same length of 7 m. Figure 9 (blue squares) presents FRD data taken for 6-MMF. As described in the Supplementary Material, each data point is the average of five measurements of output f-ratio value, and the error bars represent the minimum and maximum measured values. One can clearly observe the difference in FRD mitigation performance between the two fiber links in Fig. 9, with the PL-MCF-PL link exhibiting a significant improvement in FRD performance than 6-MMF. This result demonstrates the potential of our PL-MCF-PL based approach to FRD mitigation in fiber-fed instrumentation [32].

Clearly, we have used different experimental methods to quantify the f-ratio of the light used to excite the fiber link, and the f-ratio of the light exiting the fibre link. As a result, caution is required when directly comparing the absolute magnitudes of these two f-ratio measurements. Regardless, since our f-ratio measurement methods do not change for different fiber links, our experimental approaches are adequate and appropriate to observe qualitative differences in the FRD performance of different multimode fibre links.

To investigate how the FRD mitigation performance varies with wavelength, we repeated the measurements for wavelengths of $1424$ nm $\pm 42$ nm and $1064$ nm $\pm 25$ nm, selected by changing the bandpass filter. Figure 10(a) shows the evolution of the output f-ratio as a function of the input f-ratio for the PL-MCF-PL link, with the red circles, green diamonds, and blue triangles representing data at 1550 nm, 1424 nm and 1064 nm respectively. Again, each data point is the average of five measurements, and the error bars represent the minimum and maximum values. A full description of the output data acquisition and data processing for all the output f-ratio measurements is presented in the second section of the Supplementary Material.

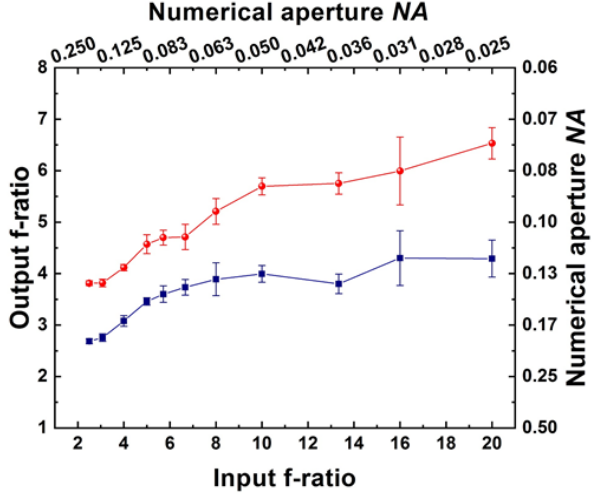

**Fig. 9**. Output f-ratio versus input f-ratio measured for the 6-MMF (blue squares) and the PL-MCF-PL link (red points), at the wavelength of 1550 ± 20 nm.

A number of interesting features can be observed in Fig. 10(a). It is clear, for example, that the minimum output f-ratio decreases with decreasing wavelength, and that the overall FRD performance of the PL-MCF-PL link degrades with deceasing wavelength. This, we believe, is primarily because the PL transitions of the PL-MCF-PL link become progressively less adiabatic as the wavelength is reduced. As a result, higher-order modes are excited at both the input and output PLs, decreasing the output f-ratio and degrading the FRD properties. Figure 10(b-c) and Fig. 10(d-e) represent similar characterization results to those presenting in Fig. 4(g-h) using 1550 nm light, but this time using light at 1424 nm and 1064 nm and either optimized (b & d) or non-optimized input excitation positions (c & e). One can clearly observe that the larger cores become progressively multimode when the wavelength is reduced. These higher-order modes can more efficiently cross-couple within the MCF than the smaller fundamental modes, further degrading the FRD performance of the fiber link. The higher-order modes supported by the cores facilitate cross-talk along the MCF, and induce FRD through the transfer of energy from lower-order modes at the input to higher-order modes at the output.

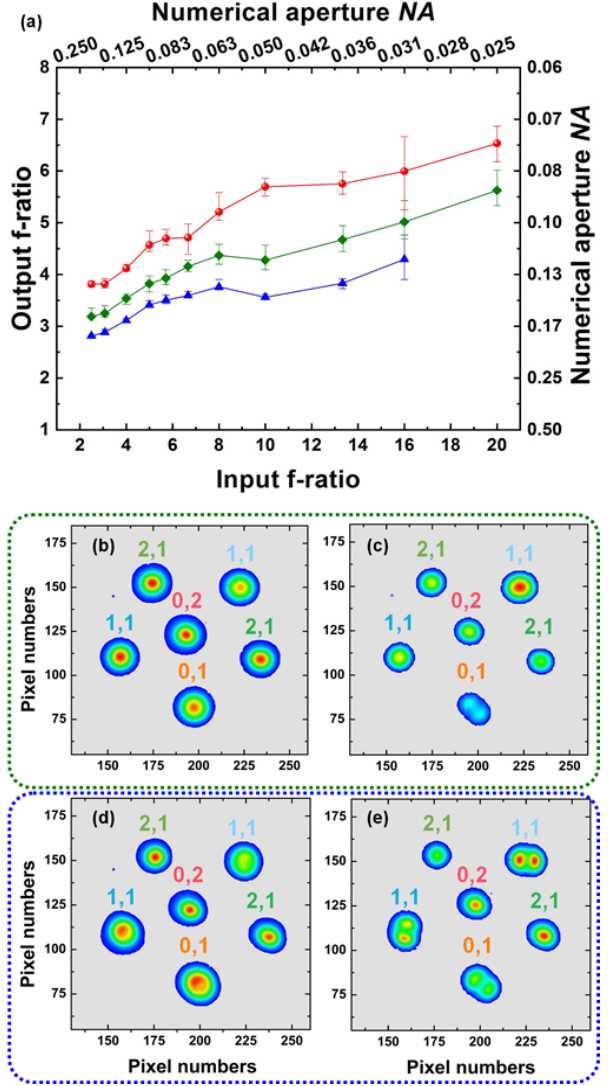

**Fig. 10.** (a) Output f-ratio versus input f-ratio measured for the PL-MCF-PL at the wavelengths of 1550 nm ± 20 nm (red circles), 1424 nm ± 42 nm (green diamonds), and 1064 nm ± 25 nm (blue triangles). (b-e) Composite near-field images of the MCF output (no PL transitions) when exciting individual MCF cores using either (b & c) 1424 nm ± 42 nm or (d & e) 1064 nm ± 25 nm light. In (b & d) the input excitation is optimized, and in (c & e) the input excitation is deliberately misaligned to excite higher-order modes.

## 5. CONCLUSION

We have reported FRD mitigation in a multimode fiber link using a MCF with mode-selective PLs. The MCF has 6 Ge-doped step-index cores with dissimilar diameters, and sufficient inter-core separations to minimize inter-core cross-talk over the length of fiber used. The FRD mitigation was achieved by fabricating mode-selective PLs at each end of the MCF. When tested at 1550 nm, the FRD behavior of the PL-MCF-PL link at 1550 nm is superior to that of a step-index MMF that also supports 6 modes. We believe this is due to the manner in which the PL-MCF-PL link inhibits the transfer of energy from low-order modes at the input PL to high-order modes at the output PL.

FRD was observed to become increasingly pronounced as the excitation wavelength was reduced from the design wavelength of 1550 nm – a behavior that we attribute to the fact that the PLs at either end of the PL-MCF-PL link become progressively less adiabatic as the wavelength is reduce, and due to the increasingly multimode nature of the MCF cores, which then facilitates core-to-core cross-talk.

In summary, we have demonstrated how MCFs and mode-selective PLs open an entirely new route to control and mitigate FRD in multimode fiber links, with considerable potential to impact the design of fiber-fed instruments in fields such as astronomy. Our approach is applicable to other regions of the spectrum, such as the visible, by fabricating a scaled version of the MCF. We also believe that the concept we have presented is applicable to fiber links operating with many 10's of modes.

**Funding.** This work was funded by the UK STFC through grants ST/N000544/1 and ST/N000625/1, and by the European Union Horizon 2020 grant 730890 (OPTICON). Raw data will be made available on the Heriot Watt University PURE system.

# Supplementary Material - Focal-ratio-degradation (FRD) mitigation in a multimode fiber link using mode-selective photonic lanterns

Aurélien Benoît,[1] Stephanos Yerolatsitis,[2] Kerrianne Harrington,[2] Tim A. Birks,[2] and Robert R. Thomson[1]

[1]*SUPA, Institute of Photonics and Quantum Sciences, Heriot-Watt University, Edinburgh, EH14 4AS, UK*
[2]*Department of Physics, University of Bath, Claverton Down, Bath BA2 7AY, UK*
*Corresponding author: a.benoit@hw.ac.uk

This document provides supplementary information relating to "Focal-ratio-degradation (FRD) mitigation in a multimode fiber link using mode-selective photonic lanterns". We present a step by step description of how the power distribution across the MCF cores was measured as a function of the f-ratio of the light used to excite the multimode port of the photonic lantern. We also provide a complete description of how the output f-ratio data was acquired and processed.

## 1. MCF POWER DISTRIBUTION AFTER THE PL

In the ideal case, a mode-selective PL couples each spatial mode of its multimode port to a specific single mode core at its output. We therefore expect that since different spatial modes of the multimode port are excited as the f-ratio of the excitation light is varied, the distribution of light across the MCF cores will also change. We characterized this and present the results in Fig. 6 of the main manuscript. Figure S1 presents the full results. The f-ratio of the excitation light is indicated in the bottom left of each sub-figure. This full sequence is presented in the visualization 1.

As can be seen in Fig. S1, the light becomes progressively more localized into fewer and fewer MCF cores as the excitation f-ratio is increased. This behavior is most pronounced when an excitation f-ratio of 20 is used. Under this case, only the fundamental mode of the multimode port of the PL is efficiently excited, which then excites only one core of the MCF at the output.

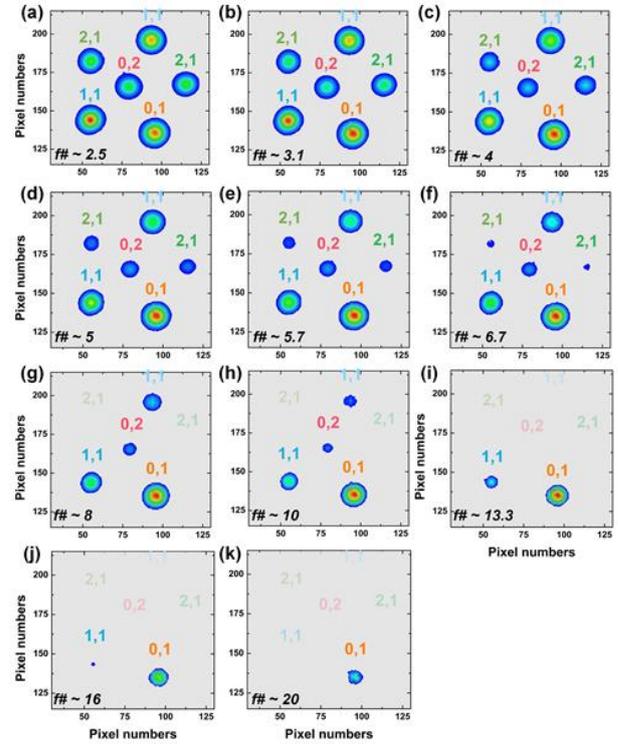

**Fig. S1.** (a-k) Evolution of the near-field intensity distributions of the output MCF as a function of the excitation f-ratio from (a) the largest pinhole diameter to (k) the smallest.

## 2. FRD DATA ACQUISITION AND PROCESSING

The aim of the experiment is to measure how the f-ratio of the light emerging from the fiber link under test varies as we adjust the f-ratio of the excitation light.

A variety of different methods have been used to characterize the focal-ratio degradation (FRD) of multimode fibers. These can involve the use of angled or non-angled input beams, and the detection of either collimated or uncollimated output light [1-4]. For our purposes, we chose to use a collimated input beam that is focused onto the input of the fiber link under test with a variable f-ratio. At the output of the fiber link under test, we measure the f-ratio of the emerging light using an InGaAs camera. The experimental setup is described in Fig. 7(b) of the main manuscript. The full data processing from the five raw image acquisitions to a final output f-ratio value follows these main steps:

### STEP 1 - Raw data
Acquisition of the five raw output beam images ($Ø_1$ to $Ø_5$ in Fig. 7(b)) from the fibre link under test with the InGaAs camera placed on a translation stage for each input excitation f-ratio (Fig. S2(STEP 1)). The first position ($Ø_1$) of the camera in front of the fibre link under test is optimized to obtain the best signal-to-noise. From $Ø_1$ to $Ø_5$, the camera is translated by 0.2 mm between each measurement. For input f-ratios beyond 10, this movement was reduced to 0.1 mm due to the reduction of the input power with a smaller adjustable graduated pinhole.

### STEP 2 - Virtual knife edge technique
To avoid assuming any specific beam shape, a virtual knife-edge technique is used on the raw images to integrate each column and obtain an S-shaped cumulative curve (Fig. S2(STEP 2)).

### STEP 3 - Determination of 95 % of the beam
A derivative function is applied to this cumulative S-curve to achieve recover a representative shape of the emergent beam which is normalized. For the smallest output beam (highest f-ratio), the beam size is just described by a few points. A linear extrapolation allows us to correctly define the encircled energy at 95 % of the full beam diameter, as shown by the blue arrow in Fig. S2(STEP 3).

### STEP 4 – Output f-ratio value
With the five beam size measurements at 95 % of the beam (black squares in Fig. S2(STEP 4)), a linear fit is applied to smooth the extrapolation and to increase the precision, giving five new beam sizes ($Ø_1$ to $Ø_5$ in green). A f-ratio value is then calculated for each comparison between two beam sizes, $Ø_1$ and $Ø_2$, $Ø_1$ and $Ø_3$ etc.

### STEP 5 - Final f-ratio value
The same process was repeated five times. All the f-ratio measurements in Figs. 9 and 10 are an average of five measurement campaigns, with the maximum and minimum values giving the error bars (Fig. S2(STEP 5)).

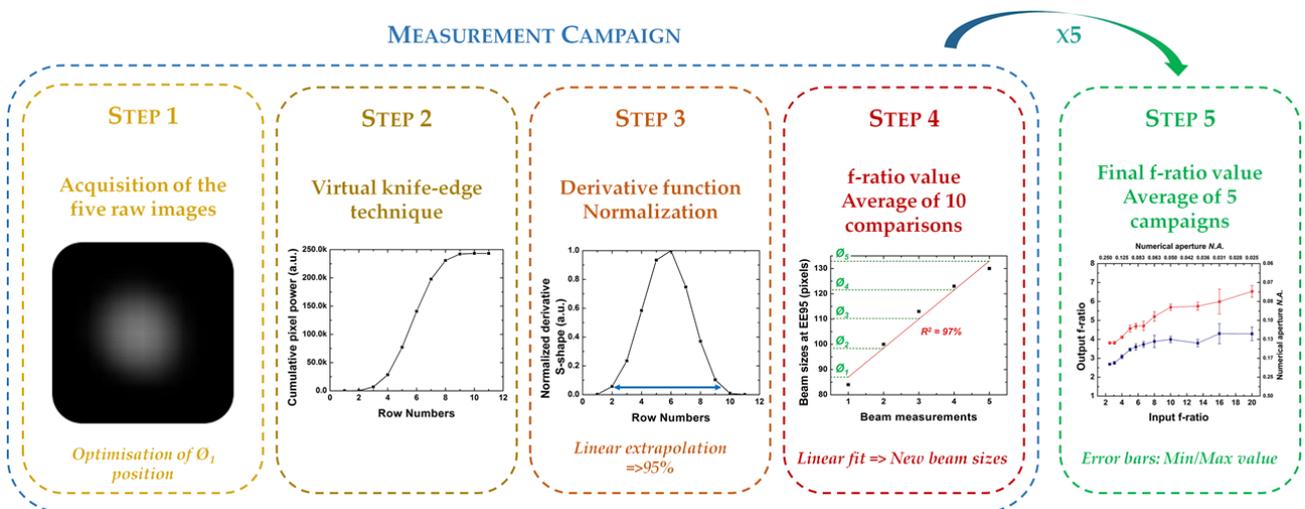

**Fig. S2.** Steps used to process the raw data acquired to measure the output f-ratio. STEP 1: the InGaAs camera is used to capture the output beams at different distances from the output of the fiber link under test. STEP 2: a virtual knife-edge technique is used to integrate the beam shown in STEP 1. STEP 3: the data in STEP 2 is then differentiated and normalized. The blue arrow represents the measurement of the encircled energy at 95 % of the beam size. STEP 4: a linear fit (red line) is applied to the five beam sizes from STEP 3 (black squares) to determinate precisely the five new beam sizes (from $Ø_1$ to $Ø_5$ in green) and calculate an output f-ratio value. STEP 5: The measurement campaign is reproduced 5 times to calculate the final f-ratio value. This figure presents an example of the data acquisition and processing for the mode-selective PL at an excited f-ratio of 13.3.